\documentclass{ifacconf}

\usepackage{amsmath}
\usepackage{graphicx}      %
\usepackage{natbib}        %
\usepackage{amssymb}
\usepackage{siunitx}
\usepackage{latexAbbreviations}
\usepackage{tabularx}
\usepackage{multirow}
\usepackage{multicol}
\usepackage{longtable,booktabs}
\usepackage{subcaption}
\usepackage{pgfplots}
\usetikzlibrary{external}
\usetikzlibrary{calc} %
\newcommand{\matlab}{\texttt{MATLAB}\ }
\newcommand{\casadi}{\texttt{CasADi}\ }
\newcommand{\ipopt}{\texttt{IPOPT}\ }
\def\centerarc[#1](#2)(#3:#4:#5){ \draw[#1] ($(#2)+({#5*cos(#3)},{#5*sin(#3)})$) arc (#3:#4:#5) }

\usepackage[absolute,showboxes]{textpos}

\TPMargin{5pt}

\newcommand{\copyrightstatement}{
    \begin{textblock}{0.87}(0.06,0.93)    
        \noindent
        \centering
        \textcopyright 2020 the authors. This work has been accepted to IFAC for publication under a Creative Commons Licence CC-BY-NC-ND
    \end{textblock}
}

\begin{document}
\begin{frontmatter}

      \title{Flatness-based MPC for underactuated surface vessels in confined areas}

      \author{Simon Helling~~}
      \author{Max Lutz~~}
      \author{Thomas Meurer}
      
      \address{Chair of Automatic Control, Kiel University, Kaiserstr. 2, 24103 Kiel, Germany (e-mail: \{sh, mlut, tm\}@tf.uni-kiel.de).}
      
      \begin{abstract}%
      	A two-phase model predictive controller (MPC) is proposed for
        underactuated surface vessel operation in confined
        environments. For general driving maneuvers (phase one) the ship's
        geometry is not considered explicitly while in more restricted
        areas (stage two) which occur, e.g., in mooring maneuvers, the ship's
        geometry is approximated to ensure collision avoidance.
        To remove the dynamical constraint in the problem setup, the
        differential flatness of the fully actuated system is
        exploited and the flat outputs are parameterized using
        B-spline functions. Underactuated behavior is retained by
        means of inequality constraints that are imposed on the
        non-controllable input. In an effort to solve the MPC, a
        static nonlinear optimization problem is formulated and
        feasibility w.r.t. obstacles and actuator constraints is
        ensured at collocation points. Static obstacles are considered
        as constructive solid geometry functions in the MPC which also
        takes into account disturbances induced by wind. 
      \end{abstract}
      
      \copyrightstatement
\end{frontmatter}

\section{Introduction}\label{sec:introduction}
Recent years show an advancing interest in the field of autonomous
vessels. This is due to the variety of challenging applications where
autonomous systems can be advantageous to humanly-operated vessels but also because
of the task to solve the arising complex problems that involve
environmental disturbances and nonlinear vessel dynamics, see \cite{Streng2020}.

Along with
classical path-following scenarios using PID controllers as shown in
\cite{Barslett2018}, more advanced approaches such as Lyapunov-based
methods involving, e.g., passivity and backstepping techniques were applied
in
\cite{Fossen2002a,Breivik2004,Do2006,Do2009a,Fossen2011}. Furthermore,
exact feedback linearization and differential flatness were exploited in
\cite{Agrawal2004a,DeAquinoLimaverdeFilho2013,Paliotta2018}. In general,
the mentioned approaches are not able to handle input and state
constraints. To deal with such issues, a third branch has
emerged which utilizes optimization-based techniques, see, e.g., \cite{Bitar2018,Bitar2019,Lekkas2016}.
Essentially, optimization-based methods aim to minimize a cost
functional depending on the control inputs subject to the system dynamics and
additional equality and inequality constraints. Methods to solve this optimal control problem (OCP) can be characterized
as indirect or direct, where the former leads to a two-point
boundary value problem and the latter directly minimizes the
cost functional by suitable discretization.

While nonlinear and optimization-based techniques constitute
independent methods, their combination can lead to
increased performance and reduced complexity. This combined approach
goes back to \cite{Agrawal1998} and was further extended to the class
of differentially flat systems, e.g. in \cite{Milam2000}. Herein, the
so-called flat outputs are parameterized with B-spline functions to
obtain an OCP, where the constraint imposed by the system dynamics is implicitly 
fulfilled. Therefore, this constraint can be omitted in the problem
setup. Subsequent discretization in time transfers the OCP to a
static optimization problem (direct method). This approach has already
been used for fuel optimization in hybrid electric drives and
trajectory generation for quadrocopters, see \cite{Abel2015} and
\cite{Abel2016}, respectively.

In this contribution, the combined flatness and optimization approach
is extended and applied to an underactuated surface vessel model. In
\cite{Agrawal2004a} the flatness of the considered model is verified
under restrictive assumptions on the model parameters. Moreover, the
resulting flat state and input parameterizations contain several
singularities, which severely restrict its applicability.
To address this in the following the so-called defect elimination
method is used as suggested, e.g., in \cite{Oldenburg2002}. Utilizing
this approach, the underactuated dynamics is achieved by means of the
singularity free flat parameterization obtained for a fully actuated
vessel model. This comes
at the cost of an additional 
equality constraint that must be
imposed on the parameterized, non-controllable input. For practical reasons, 
however, this equality constraint is replaced by two inequality constraints.
This approach is evaluated for driving maneuvers in confined environments
including mooring based on closed-loop MPC
involving disturbances induced by wind. Herein, the maneuver is separated 
into two phases. The first phase will be referred to as the driving phase
where the ship geometry is not explicitly considered to evaluate obstacle 
collisions. Subsequently, the second phase will be referred to as the
mooring phase, where the ship geometry is approximated to ensure obstacle 
avoidance for the entire ship hull. 
 
The paper is organized as follows. The vessel model is introduced in Section
\ref{sec:surface_vessel_model} together with its flat state and input
parameterization. Section \ref{sec:optimal_control} describes the general
form of an OCP and introduces the used approach for obstacle modeling with
constructive solid geometry (CSG) functions. Additionally, the
flatness-based direct solution method is described by briefly
introducing the main properties of B-spline functions and formulating
their connection to flat outputs. To account for wind-induced
disturbances, the extension to MPC is proposed in Section
\ref{sec:model_predictive_control}. Subsequently, a two-phase MPC
is presented together with short remarks on the used disturbance model
which is assumed to be unknown to the MPC. Finally, Section
\ref{sec:simulation_results} shows simulation results and the paper
closes with some conclusions in Section \ref{sec:conclusion}. 
\section{Surface vessel model}\label{sec:surface_vessel_model}
\begin{figure}[t]
	\begin{center}
		\includegraphics{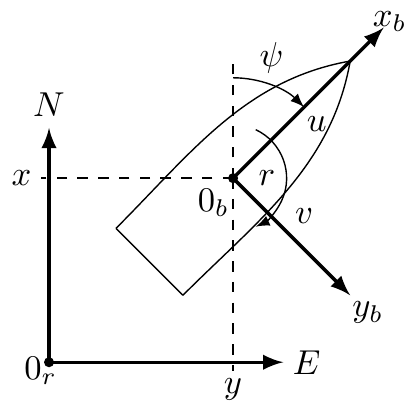}
		\caption{Vessel position and orientation in NED frame and velocities in body-fixed frame for 3DOF surface vessel.}
		\label{fig:vessel_schematic}
	\end{center}
\end{figure}
Assuming that the vessel operates in calm sea conditions, e.g., in
harbor areas or near shore 
shipping applications, roll, pitch
and heave velocities can be neglected. This results in a three degrees
of freedom (3DOF) description of a surface vessel for which two sets of
coordinates are required. The first set $\veta=\transpose{[x\ y\
  \psi]}$ describes the vessel location and pose in the
North-East-Down (NED) frame with origin $0_r$, where $x$ corresponds
to the north and $y$ to the east coordinate. The third component
$\psi$ describes the vessel orientation w.r.t. the north axis. This
set of coordinates is a reference frame for the second set of
coordinates  $\vnu=\transpose{[u\ v\ r]}$ which represents the vessels
surge and sway velocities as well as its yaw rate in a body-fixed
coordinate frame, respectively. These relations can be observed in
Fig. \ref{fig:vessel_schematic}.

\subsection{Vessel dynamics}
By applying Newton's second law the equations of motion for a surface
vessel can be described using matrix-vector notation, see
\cite{Fossen2002a}, in the form
\begin{subequations}
	\label{eq:dxdt_robot}
	\begin{align}
	\label{eq:rpsi} 
	\vetad &= R(\psi)\vnu\\
	\label{eq:dnudt} 
	M\vnud &= -\big(\cnu+\dnu\big)\vnu + B_{\tau}\vtau_{\textit{c}} + \vtau_{\text{w}}
	\end{align}
\end{subequations}
where
\begin{align}
	\rpsi = \rpsimat
\end{align}
is the rotation matrix and
\begin{align}
  M \!=\!\! 
  \begin{bmatrix}
    m_{11} & 0 & 0 \\
    0 & m_{22} & m_{23}\\
    0 & m_{32} & m_{33}
  \end{bmatrix}
                 \!\!=\!\!
                 \begin{bmatrix}
                   m\!-\!X_{\dot{u}} &0 & 0\\
                   0 & m\!-\!Y_{\dot{v}} & mx_g \!-\! Y_{\dot{r}}\\
                   0 & mx_g \!-\!N_{\dot{v}} & I_{zz} \!-\! N_{\dot{r}}\\
                 \end{bmatrix}
\end{align}
describes the mass matrix with vessel mass $m$, hydrodynamic derivatives in SNAME notation $X_{\dot{u}},Y_{\dot{v}},Y_{\dot{r}},N_{\dot{v}},N_{\dot{r}}$, distance $x_g$ of the origin $0_b$ to the center of gravity on the $x_b$-axis, and moment of inertia $I_{zz}$. Coriolis and centripetal effects are included in the matrix
\begin{align} 
	\cnu = -\transpose{\cnu}=\begin{bmatrix}
	0 & 0 & c_{13}\\
	0 & 0 & c_{23}\\
	-c_{13} & -c_{23}&0
	\end{bmatrix},
\end{align}
where
\begin{align*}
  c_{13} = -m_{22}v-\frac{m_{23}+m_{32}}{2}r,\quad
  c_{23} = m_{11}u.
\end{align*}
The damping matrix
\begin{align}
	\dnu = -\begin{bmatrix}
		X_u\!+\!X_{\abs{u}u}\abs{u} & 0 & 0\\
		0&		Y_v\!+\!Y_{\abs{v}v}\abs{v} & Y_r\\
	    0& N_v&N_r\!+\!N_{\abs{r}r}\abs{r}\\
	\end{bmatrix}
\end{align}
combines linear damping terms $X_u,Y_v,Y_r,N_v,N_r$ and nonlinear
second order modulus model terms $X_{\abs{u}u}$, $Y_{\abs{v}v}$,
$N_{\abs{r}r}$. For an underactuated surface vessel it holds that the
effect of the control input $\vtau_{\textit{c}}=\transpose{[\tau_u\
  \tau_r]}$ is applied with the actuator configuration matrix  
\begin{align}
	B_{\tau}=\begin{bmatrix}
		1 & 0\\ 0&0\\0&1
	\end{bmatrix}.
\end{align}
The vector $\vec{\tau}_{\text{w}}$ describes wind-induced
disturbances. For a compact notation, the state vector $\vx =
\transpose{[\transpose{\veta}\ \transpose{\vnu}]}\inRn{n}$, where $n=6$
is the number of states, and input vector
$\vu=\vtau_{\text{c}}\inRn{m}$, where $m=2$ is the number of inputs,
are
defined such that \eqref{eq:dxdt_robot} can be rewritten in
nonlinear input-affine form 
\begin{subequations}
\begin{align}
	\label{eq:dxdt}
	\vxd = \vfvx + B\vu + \overline{\vtau}_{\text{w}}, \qquad t>0,\quad \vxnn=\vxn
\end{align}
where
	\begin{align}
		\vfvx &= 
			\begin{bmatrix}
				\rpsi\vnu\\
				-M^{-1}\big(\cnu+\dnu\big)\vnu
			\end{bmatrix},\\
		B &= 
			\begin{bmatrix}
				\vec{0}^{(3\times m)}\\
				M^{-1}B_{\tau}
			\end{bmatrix},\\
		\overline{\vtau}_{\text{w}} &=
			\begin{bmatrix}
				\vec{0}^{(3\times 1)}\\
				M^{-1}\vtau_{\text{w}}
			\end{bmatrix}.
	\end{align}
\end{subequations}

\subsection{Differential flatness}
In the following, the differential flatness of the vessel model is shown. Theoretical background 
concerning differential flatness is provided in, e.g., \cite{Fliess1995,Rothfuss1997,Fliess1999}.
The flat parameterization of the underactuated vessel 
model shows several singularities, see \cite{Agrawal2004a}. Therefore, a fully actuated model with $\vu=\vtau^\prime_{\textit{c}}=\transpose{[\tau_u\ \tau_v\ \tau_r]}$ and $B = \transpose{[\vec{0}^{(m\times 3)}\ \transpose{(M^{-1}B^{\prime}_{\tau})}]}$ where $B^\prime_{\tau}=I^{(3\times
  3)}=\text{diag}\{1,1,1\}$ is assumed. Furthermore, the disturbance term in
\eqref{eq:dxdt} is neglected so that $\overline{\vec{\tau}}_{\textit{w}}=0$. Choosing the
flat output $\vz=\veta=\transpose{[x\ y\ \psi]}$, the states and
inputs can be differentially parametrized in the form
\begin{subequations}
  \label{eq:flatness_ship_theta_x_theta_u}
  \begin{align}
    \label{eq:flatness_ship_theta_x}
      \vx &\!=\!
            \vec{\theta}_{\vx}\big(\vz,\vzd,\hdots,\vec{z}^{(\vec{\beta}-\vec{1})}\big)
            \!=\!
      \begin{bmatrix}
        z_1\\
        z_2\\
        z_3\\
        \sin(z_3)\dz_2+\cos(z_3)\dz_1\\
        \cos(z_3)\dz_2-\sin(z_3)\dz_1 \\
        \dz_3
      \end{bmatrix}\\
    \label{eq:flatness_ship_theta_u}
    \vu &\!=\!
          \vec{\theta}_{\vu}\big(\vz,\vzd,\hdots,\vec{z}^{(\vec{\beta})}\big)
          \!=\!
          \begin{bmatrix}
            \theta_{\tau_u}\\
            \theta_{\tau_v}\\
            \theta_{\tau_r}
          \end{bmatrix},
  \end{align}  
\end{subequations}
with $\vec{\beta}=(2 \ 2 \ 2)$. The terms $\theta_{\tau_u}$,
$\theta_{\tau_v}$, and $\theta_{\tau_r}$ are provided in Appendix
\ref{app:inputparam}. It becomes apparent that no singularities arise
in \eqref{eq:flatness_ship_theta_x_theta_u}.

To recover the original underactuated vessel dynamics from the flat
parameterization of the fully actuated vessel it is necessary to
impose the constraint
\begin{align}
  \label{eq:flatness_constraint}
  \theta_{\tau_v}=0,
\end{align}
which induces an ODE in the components of $\vz$. In principle, this ODE
can be interpreted as the internal dynamics, see, e.g., the analysis
in \cite{Rothfuss1996}.
For the
considered OCP \eqref{eq:flatness_constraint} is included by means of two inequality constraints
to be fulfilled in terms of the decision variables. 

\subsection{Model parameters}\label{sec:model_parameter}
The vessel parameters are taken from \cite{Do2006} for a model ship
and are summarized in Tab. \ref{tab:vessel_param}. Therein, $L_S$ and
$W_S$ are the vessel length and width, respectively. The inputs are
constrained according to 
\begin{subequations}
	\begin{align}
	-\SI{5}{\newton}&\leq\tau_u\leq\SI{5}{\newton},\\
	-\SI{0.2}{\newton\meter}&\leq\tau_r\leq\SI{0.2}{\newton\meter}.
	\end{align}
\end{subequations}

\begin{table}[ht]
	\begin{center}
		\captionsetup{width=\textwidth}
		\caption{Vessel parameters}\label{tab:vessel_param}
		\ra{1.2}
		\begin{tabular}{l S[table-format=2.2, table-column-width=0.35cm] l S[table-format=2.1,table-number-alignment = center,table-column-width=0.35cm] l S[table-format=1.1,table-number-alignment = center,table-column-width=0.3cm] l S[table-format=2.2,table-number-alignment = center,table-column-width=0.6cm,table-align-text-post = false] l S[table-format=1.1,table-number-alignment = center,table-align-text-post = false]}
			\toprule
			\multicolumn{2}{l}{\multirow{2}{*}{Mass matrix}} & \multicolumn{4}{l}{Damping matrix} & \multicolumn{2}{l}{\multirow{2}{*}{Vessel}}\\
			& & \multicolumn{2}{l}{linear} & \multicolumn{2}{l}{nonlinear} &  &  \\
			\midrule
			$M_{11}$&25.80&$X_u$   & -12.0&$X_{|u|u}$&-2.1&$L_S$  &1.20\ \si{\meter}&\\
			$M_{22}$&33.80&$Y_{v}$&-17.0 &$Y_{|v|v}$&-4.5&$W_S$&0.35\ \si{\meter}&\\
			$M_{23}$&6.20  &$Y_{r}$ &-0.2   &$N_{|r|r}$ &-0.1&$m$     &17.00\ \si{\kilogram}\\
			$M_{32}$&6.20  &$N_{v}$&-0.5   &                &       &            &\\
			$M_{33}$&2.76&$N_{r}$ &-0.5   &                &      &            &\\
			\bottomrule
		\end{tabular}
	\end{center}
\end{table}

\section{Flatness-based optimal control}\label{sec:optimal_control}
The aim for the desired approach is to generate trajectories 
while also considering actuator constraints. In other words, a combined
trajectory-generation and motion control of the vessel is required while also
taking into account confined environments for mooring maneuvers.
In the following, CSG functions are discussed which can represent 
arbitrary shapes. These can be included to an OCP formulation.
Furthermore, a flatness-based solution method for the OCP using B-splines 
is discussed.

\subsection{Obstacle modeling}
For obstacles of arbitrary shapes, CSG functions are used, see
\cite{Ricci1972a}. These are based on geometric primitive functions
$f^{\text{pr}}(\vx)$ such as ellipsoids, lines, and triangles. In
order to describe the surface $S$ of a
shape mathematically, a function of the form 
\begin{align}
	\label{eq:csg_inequality}
	f^{{S}}(\vx)\leq 1
\end{align} 
can be formulated which combines several primitive shapes
using the maximum operator, i.e.
\begin{align}
	f^{{S}}(\vx) = \max\left\{f^{\text{pr}}_{1}(\vx),\hdots,f^{\text{pr}}_l(\vx)\right\},
\end{align}
where $l$ is the number of primitive functions used to define the shape.
Since the gradient of the maximum operator is not smooth
the approximation
\begin{align}
	\begin{split}
		\max\{f^{\text{pr}}_1(\vx),\hdots,f^{\text{pr}}_l(\vx)\} \approx \big[&(f^{\text{pr}}_1(\vx))^p+\hdots\\
		&+(f^{\text{pr}}_l(\vx))^p  \big]^{\frac{1}{p}}
	\end{split}
\end{align}
is used, where the approximation quality increases with increasing
$p\in\mathbb{N}$. In the following scenarios, rectangles are used
to reflect confined areas. A rectangle can be constructed
from two shifted and rotated parabolas, so that 
\begin{align}
  \label{eq:csg_rectangle}
  \begin{split}
    f_{\text{rect}}^{{S}}(\vx \vert \vec{r}) &= \Bigg[\bigg(\frac{\cos(\alpha)(x-\tilde{x}_{0}) + \sin(\alpha)(y-\tilde{y}_{0})}{d_x} \bigg)^{2p}\\
    &+\bigg(\frac{-\sin(\alpha)(x-\tilde{x}_{0})+\cos(\alpha)(y-\tilde{y}_{0})}{d_y} \bigg)^{2p}\Bigg]^{\frac{1}{p}},
  \end{split}
\end{align}
where the elements of $\vec{r}=\transpose{[\tilde{x}_0\ \tilde{y}_0\
  d_x \ d_y \ \alpha \ p]}$ describe the center position, length,
width, orientation, and approximation quality parameter in the
reference frame.
\subsection{Problem formulation}
In the following, the OCP for the considered system is expressed with
\begin{subequations}
	\label{eq:ocp}
	\begin{align}
	\label{eq:ocp_J}
	&\min_{\vu} \ J(\vu) = \varphi(\tf,\vx(\tf))\\
	\notag\text{s.t.}\\
	\label{eq:ocp_dynamic_constraint}
	&\vxd = \vfvx + B\vu, \quad t>0, \quad 	\vx(0) = \vxn\\
	\label{eq:ocp_final_constraint}
	&\vec{g}\big(\tf,\vx(\tf)\big)=\vec{0}\\
	\label{eq:ocp_h}
	&\vec{h}(\vx)\leq\vec{0} \\
	\label{eq:ocp_input_constraints}
	&\vu^{-}	\leq \vu \leq \vu^{+},
	\end{align}
\end{subequations}
where $J(\vu)$ represents the cost functional in Mayer form that is to be minimized,
$\tf$ is the final time, \eqref{eq:ocp_dynamic_constraint} denotes the
ODE constraint imposed by the system dynamics with initial
condition $\vxnn=\vxn$. Furthermore, terminal path constraints are included with
\eqref{eq:ocp_final_constraint}, and state constraints imposed by obstacles are
formulated with \eqref{eq:ocp_h}. Herein, $\vh(\vx)$ is obtained by rearranging 
\eqref{eq:csg_inequality} and including \eqref{eq:csg_rectangle} which yields $h_i(\vx)=1-f_{\text{rect},i}^S(\vx), i=1,\hdots,q$, where $q$ is the number of rectangular obstacles. Input constraints are expressed using \eqref{eq:ocp_input_constraints}, where $\vu^-$, and $\vu^+$ denote the lower and upper input bounds, respectively.

\subsection{Flatness-based solution using B-splines}
The ODE constraint \eqref{eq:ocp_dynamic_constraint} is
implicitly fullfilled by the flat parameterization
\eqref{eq:flatness_ship_theta_x_theta_u} of the system. Therefore, the differential flatness
of the vessel system can be exploited when the OCP is formulated in
flat coordinates thereby eliminating the ODE constraint. Since
the problem is still an infinite-dimensional it is convenient to
parameterize the flat outputs using B-spline functions which are unions of curve segements. For this,
consider the expansion
\begin{align}
	\label{eq:z_approx_bsplines}
	z_j(t) \approx \hat{z}_j(t,\vec{p}_j) =  \sum_{i=0}^{N_j}B_{i,D_j}(t)p_{i,j},\ \
		\begin{split}
			t&\in[0,\tf],\\
		  j&=1,\hdots,m
		\end{split} 
\end{align}
for the $j$th component of the flat output $\vz$. Herein, 
$B_{i,D_j}(t)$ are basis functions of order $D_j$ and the vector
{$\vec{p}_j = \transpose{[p_{0,j}\ \hdots \ p_{N_j,j}]}$} summarizes
the individual $N_j$ weights. In general, the ability to approximate
complex function behavior is improved as $N_j$ is increased. Using
B-spline functions the basis functions can be calculated recursively using the
Cox-DeBoor scheme, see \cite{Piegl2013}, i.e. 
\begin{subequations}
	\begin{align}
		\label{eq:bsplines_coxdeboor_Bi0}
		B_{i,0}(t) &= 
			\begin{cases}
			1, \qquad \text{for } t \in [u_i,u_{i+1})\\
			0, \qquad \text{else}
			\end{cases},\\
		\begin{split}
			B_{i,j}(t) 	&= \frac{t - u_i}{u_{i+j} - u_i}B_{i,j-1}(t) \\
			&\quad+ \frac{u_{i+j+1}-t}{u_{i+j+1}-u_{i+j}}B_{i+1,j-1}(t).
		\end{split}
	\end{align}
\end{subequations}
In the recursion formula it can be seen that the time horizon
$t\in[0,\tf]$ is separated using a so-called knot vector
\begin{equation}
	\begin{split}
		\hat{\vec{u}}_j&=\transpose{[u_{0,j}\ \hdots \ u_{M_j,j}]}\quad j=1,\hdots,m.
	\end{split}
\end{equation}
At the knot points, the curve segments are joined to form the B-spline function. As can be seen from the recursion, the $i$th basis function $B_{i,D_j}(t)$ that is weighted with $p_{i,j}$ for the $j$th flat output is nonzero on the interval $t\in[u_{i,j},u_{D_j+1,j})$. Thus, choosing
\begin{align}
\hat{\vec{u}}_j= \transpose{[\underbrace{0\ \hdots \ 0}_{D_j}\ 0 \hdots\ \tf \ \underbrace{\tf\ \hdots \ \tf}_{D_j}]},
\end{align} 
results in $B_{k,0}(0)=0$ for $k<D_j$ and only $B_{D_j,0}(0)=1$, so that
\begin{align}
	\hat{z}_j(0,\vec{p}_j) &= p_{0,j}.
\end{align}
Similarly, this choice of the knot vector yields $\hat{z}_j(\tf,\vec{p}_j)= p_{N_j,j}$. In this way, 
initial and final values (of the flat outputs) are parameterized using the control points $p_{0,j}$ and $p_{N_j,j}$, respectively.
The parameter $M_j$ in $\hat{\vec{u}}$ can be determined with $M_j= D_j+N_j+1$.
The flat parameterization requires derivatives of the flat outputs up
to order $\vec{\beta}$. The $k$th order derivative of a B-spline
function is given by 
\begin{align}
\label{eq:z_dot_approx_bsplines}
\hat{z}_j^{(k)}(t,\vec{p}_j) &=\sum_{i=0}^{N_j}B_{i,D_j}^{(k)}(t)p_{i,j},\ \ 
		\begin{split}
			t&\in[0,\tf], \\
			 j&=1,\hdots,m,
		\end{split} 
\end{align}
where
\begin{equation}
	\begin{split}
		&B_{i,l}^{(k)}(t) = \frac{l}{u_{i+l}-u_i}B_{i,l-1}^{(k-1)}(t) \\
		&-\frac{l}{u_{i+l+1}-u_{i+1}}B^{(k-1)}_{i+1,l-1}(t), \quad \begin{split}k&=1,\hdots,D_j-1\\l&=0,\hdots,D_j\end{split}.
	\end{split}
\end{equation}
This means that the derivative of a B-spline function is again a B-spline
function but of lower degree. Each B-spline function is $D_j-2$ times continuously
differentiable. To avoid numerical difficulties, $D_j$ 
should be chosen as small as possible, i.e. $D_j = \beta_j + 2$. For
further properties of B-spline functions, see
\cite{Piegl2013}. Substituting
\eqref{eq:z_approx_bsplines}, \eqref{eq:z_dot_approx_bsplines} together
with \eqref{eq:flatness_ship_theta_x_theta_u} into the OCP formulation
\eqref{eq:ocp} yields an equivalent problem with the new (constant) decision
variables
\begin{align}
\overline{\vec{p}} = 
\transpose{\begin{bmatrix}
	\transpose{\vec{p}}_1  & \hdots & \transpose{\vec{p}}_m
	\end{bmatrix}}  \inRn{n_p},
\end{align}
where $n_p=\sum_{j=1}^{m}N_j$
is the number of decision variables.
Feasibility w.r.t. obstacle and input constraints \eqref{eq:ocp_h} and
\eqref{eq:ocp_input_constraints}, respectively, is checked at
collocation points, $t_k = kh,\ k=0,\hdots,N$, where $N+1$ is the
number of collocation points and $t_0=0,\ t_N=\tf$. Consequently, a
NLP is obtained.

\section{Model predictive control}\label{sec:model_predictive_control}

In the following, the flatness-based OCP approach is extended to a MPC
to compensate for wind-induced disturbances. This is done by
repeatedly solving OCPs at discrete points in time with a step time of
\mbox{$\Delta t=t_{\text{MPC}}=\text{const.}$} As a scenario, a
combined driving  and mooring maneuver is considered, each resulting
in a different OCP formulation.

\subsection{Driving phase}
In the first phase, the distance to a desired terminal position $(x_{\text{f}},y_{\text{f}})$ is minimized within the fixed MPC time horizon $\tf=t_{\text{hor}}$, i.e.,
\begin{align}
	J(\vu) = \varphi(\tf,\vx(\tf)) = (x(t_{\text{f}})-x_{\text{f}})^2 + (y(t_{\text{f}})-y_{\text{f}})^2,
\end{align} 
with
\begin{align}
	\vec{g}(\tf,\vx(\tf))={\emptyset},
\end{align}
such that no terminal condition is imposed on the problem. 
In this way, the {closest point} w.r.t. the terminal position is the
solution to the OCP. It can be assumed that while driving no confined
areas are passed by the vessel so that it is sufficient to adduce the origin $0_b$
of the body-fixed frame, i.e. $(x,y)$, in order to evaluate the obstacle
functions \eqref{eq:ocp_h}.
\subsection{Mooring phase}
If the vessel origin is within a defined radius $R_{\text{s}}$
(switching point) of the desired terminal position after an arbitrary
 iteration,
the cost functional is altered to minimize the transition time, i.e. 
\begin{align}
	J(\vu) = \varphi(\tf,\vx(\tf)) =\tf.
\end{align}
This requires the formulation of a terminal condition
\begin{align}
	\vec{g}(\tf,\vx(\tf))=\vx(\tf)-\vx_{\text{f}},
\end{align}
where $\vx_{\text{f}}$ is the arbitrary but fixed final state.
In this phase, the vessel geometry is approximated as a rectangle and
feasibility w.r.t. obstacles is ensured using four edge points of the
rectangle.

\subsection{Wind-induced disturbances}
The disturbances induced by wind $\vec{\tau}_{\text{w}}$ or
$\overline{\vec{\tau}}_{\text{w}}$, respectively, are calculated
according to \cite{Fossen2011} using a normally distributed wind
direction $\beta_w \sim \mathcal{N}(\mu_{\beta},\sigma_{\beta})$ and
an absolute wind velocity $V_{w,\text{abs}}\sim\mathcal{W}(k_{V},
\lambda_{V})$, where $k_V$ and $\lambda_V$ are shape and scale
parameters of the Weibull distribution. With this, the forces and
torque applied to the vessel can be calculated with
\begin{align}
	\vec{\tau}_{\text{w}} = \frac{1}{2}\rho \big(V_{w,\,\text{rel}}\big)^2
	\begin{bmatrix}
	C_XA_f \\
	C_Y A_l\\
	C_NA_lL_S
	\end{bmatrix},
\end{align}
where $\rho$ is the air density, $V_{w,\text{rel}}$ is the relative
wind velocity which, together with the coefficients $C_X,C_Y$, and
$C_N$, depends on the absolute wind direction $\beta_w$ and speed $V_{w,\text{abs}}$. The
parameters $L_S,A_f$ and $A_l$ are vessel length, projected frontal
and lateral areas, respectively.

\section{Simulation results}\label{sec:simulation_results}
\begin{figure}[tb]
	\begin{subfigure}{.5\textwidth}
		\includegraphics{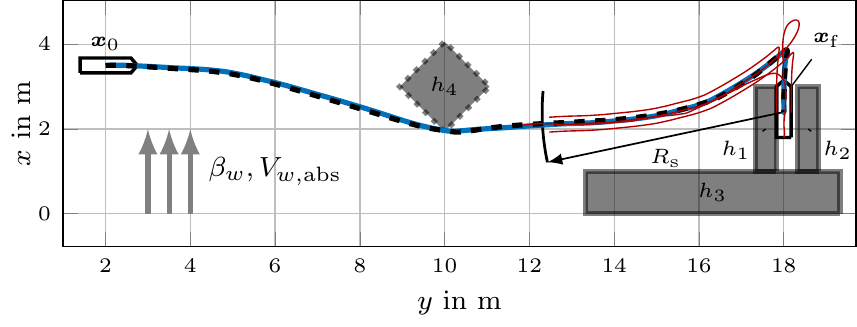}
		\captionsetup{width=.885\textwidth}
		\subcaption{Simulated path with wind direction $\beta_w$, absolute wind speed $V_{w,\text{abs}}$, switching radius $R_{\text{d}}$, initial and final positions, and obstacles $h_i(\vx),i=1,\hdots,4$, as well as edge point paths in driving phase (red).}
		\label{fig:mpc_results_xy}
	\end{subfigure}
	\begin{subfigure}{.5\textwidth}
		\begin{minipage}{.5\textwidth}
			\includegraphics{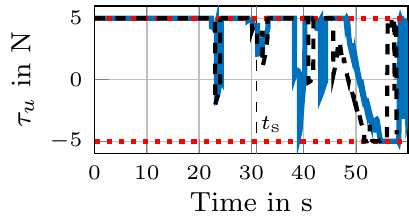}
		\end{minipage}%
		\begin{minipage}{.5\textwidth}
			\includegraphics{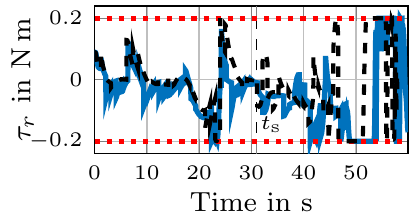}
		\end{minipage}
		\captionsetup{width=.885\textwidth}
		\subcaption{Inputs surge force and yaw torque with constraints (dashed red). The input $\tau_v$ is not explicitly shown here because it is forced to zero.}\label{fig:mpc_results_taus}
	\end{subfigure}
	\begin{subfigure}{.5\textwidth}
		\begin{minipage}{.5\textwidth}
			\includegraphics{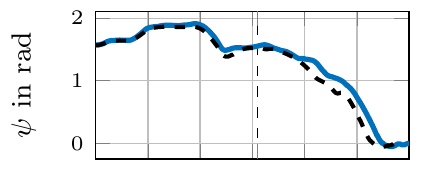}\\
			\includegraphics{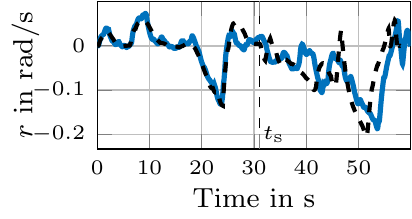}
		\end{minipage}%
		\begin{minipage}{.5\textwidth}
			\includegraphics{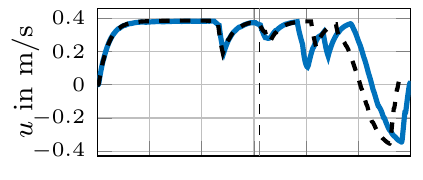}\\
			\includegraphics{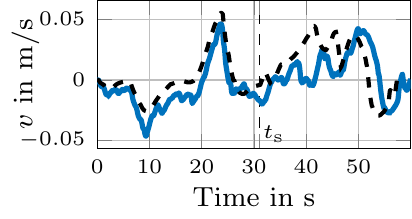}
		\end{minipage}
		\captionsetup{width=.885\textwidth}
		\subcaption{Orientation and velocities of the vessel.}\label{fig:mpc_results_states}
	\end{subfigure}
	\caption{Simulation results with optimal path (top), inputs (middle), and states (bottom) each with (blue) and without (black dotted) disturbances considering four rectangular obstacles.}
	\label{fig:mpc_results}
      \end{figure}

Simulation results are generated in \matlab using \casadi with \ipopt
as NLP solver, see \cite{Andersson2018} and \cite{Waechter2002},
respectively. The underactuated vessel dynamics using the flat
parameterization of the fully actuated system is retained by taking
into account \eqref{eq:flatness_constraint} which for numerical
purposes is approximated by   
\begin{align}
	-\epsilon \leq \theta_{\tau_v} \leq \epsilon,
\end{align}
for $\epsilon\ll 1$. For the simulation, only the solutions of $\theta_{\tau_u}$,
and $\theta_{\tau_r}$ are applied to the underactuated model.
\begin{rem}
	Setting $\epsilon=0$ would result in $N+1$ equality constraints which reduces the number of free decision variables in the NLP potentially rendering it unsolvable. Choosing $\epsilon>0$ avoids this issue.
\end{rem}
Initial and terminal (desired) states are chosen to be
\begin{subequations}
	\begin{align}
	\vxn           &= \transpose{\left[3.5\ 2\ \frac{\pi}{2}\ 0 \ 0 \ 0 \right]},\\	
	\vx_{\text{f}} &= \transpose{\left[2.4\ 18\ 0\ 0 \ 0 \ 0 \right]}.
	\end{align}
\end{subequations}
Further, the switching point is chosen to be
\begin{align}
	R_{\text{s}}=t_{\text{hor}}\sqrt{u_{\text{max}}^2+v_{\text{max}}^2},	
\end{align}
where
$u_{\text{max}}=\SI[per-mode=symbol]{0.38}{\meter\per\second},v_{\text{max}}\approx\SI[per-mode=symbol]{0}{\meter\per\second}$
describe the maximum surge and sway velocity of the vessel, respectively. The fixed time horizon is set to
$t_{\text{hor}}=\SI{15}{\second}$ in the driving phase. The MPC
horizon is shifted each iteration for
$t_{\text{MPC}}=\SI{1}{\second}$. Additionally, four obstacles are
considered where $h_i(\vx),i=1,2,3$ are relevant for the mooring
maneuver and $h_4(\vx)$ affects the driving maneuver. Feasibility
w.r.t. constraints is ensured at $N+1=200$ collocation
points. Additional scenario parameters are summarized in
Tab. \ref{tab:mpc_scenario_parameter}. The top view of the path, orientation,
initial and final position, as well as the switching point are shown in
Fig. \ref{fig:mpc_results_xy}.
It can be seen that there is no
collision with any obstacle. Figure \ref{fig:mpc_results_taus} shows
the inputs with constraints marked using dashed-red lines which are
satisfied for all times. The remainder of states is shown in
Fig. \ref{fig:mpc_results_states} together with the switching time
$t_\textit{s}=\SI{31}{\second}$. Sudden changes in the inputs can be
explained by numerical issues and disturbances which could push the
vessel into the obstacles resulting in feasibility issues for the NLP 
solver. This could be avoided using soft constraints as described in
\cite{Scokaert1999a}.
\begin{table}[tb]
	\begin{center}
		\captionsetup{width=.5\textwidth}
		\caption{Obstacle and wind parameters.}
		\label{tab:mpc_scenario_parameter}
		\ra{1.2}
		\begin{tabularx}{.48\textwidth}{l p{.5cm} p{.5cm} p{.5cm} p{.5cm} X l X l}
			\toprule
			\multicolumn{5}{c}{Obstacles} &\multicolumn{4}{c}{\multirow{2}{*}{Wind}}\\
			&$\vec{r}_1$&$\vec{r}_2$ &$\vec{r}_3$ &$\vec{r}_4$&&&&\\
			\midrule
			$\tilde{x}_0$ & 2 & 2 & 0.5 &3&$A_f$&\SI{0.35}{\square\meter}&$\mu_{\beta}$&\SI{0}{\radian}\\
			$\tilde{y}_0$ & 17.575 & 18.575&16.325& 10 &$A_l$&\SI{1.2}{\square\meter}&$\sigma_{\beta}$&\SI{0.06}{\radian}\\
			$d_x$  & 2  & 2 & 1 &1.5 & $L_S$ & \SI{1.2}{\meter} & $\lambda_V$&{0.194}\\
			$d_y$  & 0.5  & 0.5 & 6 &1.5&&& $k_V$&2\\
			$\alpha$  & 0 & 0 & 0 &{$\frac{\pi}{4}$}&&&$\rho$&\SI[per-mode=fraction]{1.205}{\kilogram\per\cubic\meter}\\
			$p$	& 12 & 12 & 12&12 & &&&\\
			\bottomrule
		\end{tabularx}
	\end{center}		
\end{table}
      
\section{Conclusion}\label{sec:conclusion}
In this paper a flatness-based MPC for an underactuated
nonlinear surface vessel model is introduced. The {fully} actuated system is
shown to be differentially flat so that the ODE constraint in
the OCP can be removed. The flat outputs are parameterized using
B-spline functions. A discretization in time of the OCP in flat
coordinates allows the formulation of a NLP which can be solved
numerically. Underactuated vessel dynamics are retained using
inequality constraints imposed on the non-controllable
input and obstacles are included to the OCP using CSG functions which
can approximate arbitrary shapes. The concept is evaluated in a
two-phase simulation scenario resulting in different OCP
formulations. Future work focuses on real-time feasibility which can
be achieved by approximating the highest-order derivative of each flat
output and subsequent integration thus avoiding recursive computation of
basis functions as shown in  \cite{Oldenburg2002}. Further work also
focuses on soft constraints and extending the concept to include
collision avoidance regulations (COLREGS).

\appendix
\section{Input parametrization}\label{app:inputparam}
The terms arising in \eqref{eq:flatness_ship_theta_u} read
\begin{subequations}
	\begin{align}
		\begin{split}
			\theta_{\tau_u} &= -X_{|u|u}\big(\sin(z_3)\dz_2+\cos(z_3)\dz_1\big)\\
			&\quad \cdot \abs{\sin(z_3)\dz_2+\cos(z_3)\dz_1}+m_{11}\sin(z_3)\ddot{z}_2\\
			&\quad -\big[(m_{22}-m_{11})\cos(z_3)\dz_3+X_u\sin(z_3)\big]\dz_2\\
			&\quad -\big[(m_{11}-m_{22})\sin(z_3)\dz_3+X_u\cos(z_3)\big]\dz_1\\
			&\quad  -\frac{1}{2}(m_{23}+m_{32})\dz_3^2+m_{11}\cos(z_3)\ddot{z}_1,
		\end{split}\\
		\begin{split}
			\theta_{\tau_v} &= Y_{|v|v}\big(\sin(z_3)\dz_1-\cos(z_3)\dz_2\big)\\
			&\quad \cdot \abs{\cos(z_3)\dz_2-\sin(z_3)\dz_1}+M_{22}\cos(z_3)\ddot{z}_2\\
			&\quad+\big[(m_{11}-m_{22})\sin(z_3)\dz_3-Y_v\cos(z_3)\big]\dz_2\\
			&\quad+\big[(m_{11}-m_{22})\cos(z_3)\dz_3+Y_v\sin(z_3)\big]\dz_1\\
			&\quad+m_{23}\ddot{z}_3-Y_r\dz_3-m_{22}\sin(z_3)\ddot{z}_1,
		\end{split}\\
		\begin{split}
			\theta_{\tau_r} &=m_{32}\cos(z_3)\ddot{z}_2+m_{33}\ddot{z}_3-N_{|r|r}\dz_3\abs{\dz_3}-N_r\dz_3\\
			&\quad+\big[(m_{22}-m_{11})\sin(z_3)\cos(z_3)\big]\dz_2^2\\
			&\quad  +\big[\big((m_{11}-m_{22})(\sin^2(z_3)-\cos^2(z_3))\big)\dz_1\\
			&\quad +\frac{1}{2}(m_{23}-m_{32})\sin(z_3)\dz_3-N_v\cos(z_3)\big]\dz_2\\
			&\quad +\big[(m_{11}-m_{22})\sin(z_3)\cos(z_3)\big]\dz_1^2\\
			&\quad +\big[\frac{1}{2}(m_{23}-m_{32})\cos(z_3)\dz_3+N_v\sin(z_3)\big]\dz_1\\
			&\quad -m_{32}\sin(z_3)\ddot{z}_1.
                      \end{split}
	\end{align}
      \end{subequations}

\bibliography{refs}             %

\end{document}